\newcommand{\msbar}{{\overline {\rm MS}}}
\def\lsim{\raise0.3ex\hbox{$<$\kern-0.75em\raise-1.1ex\hbox{$\sim$}}}
\def\gsim{\raise0.3ex\hbox{$>$\kern-0.75em\raise-1.1ex\hbox{$\sim$}}}
\def\simgt{\rlap{\lower 6.0 pt\hbox{$\mathchar \sim$}}\raise 2.5pt \hbox {$>$}}
\def\simlt{\rlap{\lower 6.0 pt\hbox{$\mathchar \sim$}}\raise 2.5pt \hbox {$<$}}
\documentclass{PoS}

\title{2+1 flavor lattice QCD simulation with $O(a)$-improved Wilson quarks }

\ShortTitle{2+1 flaavor lattice QCD simulation with $O(a)$-improved Wilson quarks }

\author{PACS-CS Collaboration : 
\speaker{N. Ukita}${}^{a}$\thanks{E-mail: ukita@ccs.tsukuba.ac.jp},
 S.~Aoki${}^{b,c}$,
 N. Ishii${}^{a}$,
 K.-I.~Ishikawa${}^{d}$,
 N.~Ishizuka${}^{a,b}$,
 T. Izubuchi${}^{c,e}$,
 D. Kadoh${}^{a}$,
 K.~Kanaya${}^{b}$,
Y. Kuramashi${}^{a,b}$,
Y. Namekawa${}^{a}$,
 M.~Okawa${}^{d}$,
 Y.~Taniguchi${}^{a,b}$,
 A.~Ukawa${}^{a,b}$,
 T.~Yoshi\'e${}^{a,b}$
 \\
 \llap{${}^a$}Center for Computational Sciences, University of Tsukuba, Tsukuba, Ibaraki 305-8577, Japan\\
 \llap{${}^b$}Graduate School of Pure and Applied Sciences, University of Tsukuba, Tsukuba, Ibaraki 305-8571, Japan\\
 \llap{${}^c$}Riken BNL Research Center, Brook-haven National Laboratory, Upton, New York 11973, USA\\
 \llap{${}^d$}Graduate School of Sciences, Hiroshima University, Higashi-Hiroshima, Hiroshima 739-8526, Japan\\
 \llap{${}^e$}Institute for Theoretical Physics, Kanazawa University, Kanazawa, Ishikawa 920-1192, Japan}

\abstract{We present simulation details and results for 
the light hadron spectrum in $N_f = 2 + 1$ lattice QCD with
the nonperturbatively $O(a)$-improved Wilson quark action 
and the Iwasaki gauge action. 
Simulations are carried out at a lattice spacing of 0.09 fm 
on a (2.9fm$)^3$ box using the PACS-CS computer. We 
employ the L\"uscher's domain-decomposed HMC algorithm with several 
improvements 
to reduce the degenerate up-down quark mass toward the physical value. 
So far the  resulting pseudoscalar meson 
mass is ranging from 702MeV down to 156MeV. We discuss on the
stability and the efficiency of the algorithm.  
The light harden spectrum extrapolated 
at the physical point is compared with the experimental 
values. We also present the values of the quark masses and the 
pseudoscalar meson decay constants.}

\FullConference{The XXVI International Symposium on Lattice Field Theory\\
		 July 14-19 2008\\
		 Williamsburg, Virginia, USA}

\begin{document}

\section{Introduction}
The PACS-CS (Parallel Array Computer System for Computational Sciences) 
project\cite{aoki,ukawa,kuramashi,kuramashi2,ukita,kadoh2,kuramashi3,pacscs}
aims at $N_f=2+1$ lattice QCD calculations at the physical point 
to remove the most troublesome systematic errors associated with 
the chiral extrapolations. 
So far our simulation points cover from 67MeV to 3.5MeV for the
degenerate up-down quark mass with the strange quark mass fixed  around
the physical value.
The reduction of $m_{\rm ud}$ down to 10 MeV is achieved
by the domain-decomposed Hybrid Monte Carlo (DDHMC) algorithm 
with the replay trick\cite{luscher,kennedy}. 
For the simulation at $m_{\rm ud}=3.5$ MeV we incorporate some 
algorithmic improvements such as 
the mass preconditioning\cite{hasenbusch,hasenbusch2}, the chronological 
inverter\cite{brower} and the deflation teqnique\cite{parks} 
which make simulations stable and contribute to reduce 
the simulation cost. For the strange quark part we employ 
the UV-filtered Polynomial Hybrid Monte Carlo (UVPHMC) 
algorithm\cite{ishikawa}. 

In this report we present the simulation details and some eminent 
results for the hadron spectrum. Chiral analyses on  
the pseudoscalar meson sector with the SU(2) and SU(3) 
chiral perturbation theories 
and calculation of the charm quark systems with the relativistic 
heavy quark action are given in separate reports\cite{kadoh, namekawa}. 

\section{Simulation details}
We employ the $O(a)$-improved Wison quark action with 
a nonperturbative improvement coefficient $c_{\rm sw}=1.715$\cite{csw} 
and the Iwasaki gauge action\cite{iwasaki}. All the 
simulations are carried out  on a $32^3\times64$ lattice 
at $\beta=1.90$ corresponding to the lattice spacing of $a=0.09$ fm. 
Table~\ref{tab:param} summarizes our simulation parameters. 
We choose combinations of the hopping parameters 
$(\kappa_{\rm ud},\kappa_{\rm s})$ based on 
the previous CP-PACS/JLQCD results\cite{cppacs/jlqcd1,cppacs/jlqcd2} 
except $(\kappa_{\rm ud},\kappa_{\rm s})=(0.137785,0.13660)$ 
which is adjusted at the physical point with the use of the PACS-CS results 
in an early stage\cite{pacscs,kuramashi2}. The physics results at  
$(\kappa_{\rm ud},\kappa_{\rm s})=(0.137785,0.13660)$ is presented 
in Ref.~\cite{kuramashi3}.
 
The DDHMC algorithm is implemented for the up-down quark  
by domain-decomposing the full lattice with an $8^4$ block size 
as a preconditioner for HMC. The domain-decomposition factorizes 
the up-down quark determinant into the UV and the IR parts
geometrically. 
As a result we have the gauge force and the up-down quark force 
with the UV and the IR parts
in the molecular dynamics evolution.  The reduction of the simulation
cost is achieved by applying the multiple time scale
integrator\cite{sexton} to these 
three forces. We find that the relative magnitude
of the force terms is given as follows:
\begin{eqnarray}   
||F_{\rm G}||:||F_{\rm UV}||:||F_{\rm IR}|| \approx 16:4:1,
\end{eqnarray}
where $F_{\rm G}$ denotes the gauge part and $F_{\rm UV, IR}$ for the UV
and the IR parts of the up-down quark. The associated step
sizes 
$\delta\tau_{\rm G}, \delta\tau_{\rm UV}, \delta\tau_{\rm IR}$ are
chosen such that
\begin{eqnarray}
 \delta\tau_{\rm G} ||F_{\rm G}|| \approx \delta\tau_{\rm UV} ||F_{\rm UV}|| \approx \delta\tau_{\rm IR} ||F_{\rm IR}||.
 \end{eqnarray} 
These step sizes are controlled by three integers $N_0, N_1, N_2$ as 
 $\delta\tau_{\rm G}=\tau/N_0 N_1 N_2, \delta\tau_{\rm UV}=\tau/N_1
N_2,$ $\delta\tau_{\rm IR}=\tau/N_2$ with $\tau$ the trajectory length.
We fix $N_0=N_1=4$ in our all simulations. The value of $N_2$ is 
adjusted to make the simulation stable.
The threshold for the replay trick is chosen to be $dH=2$.
For the strange quark we employ the UVPHMC algorithm, where the domain-deconposition is not used.
The polynomial order $N_{\rm poly}$ for the UVPHMC algorithm 
is adjusted to keep high acceptance rate for the global Metropolis test 
at the end of each trajectory. 
Based on our observation of $||F_{\rm s}||\approx ||F_{\rm IR}||$ 
for the strange quark force, 
we set $\delta\tau_{\rm s}=\delta\tau_{\rm IR}$.
Calclation of the IR force requires 
the inversion of the Wilson-Dirac operator on
the full lattice, which is carried out by the SAP (Schwarz alternative
procedure) preconditioned GCR algorithm. We use
the SSOR preconditioned GCR algorithm for the UV part. 
These preconditionings are accelerated with the single precision
arithmetic. We employ the stopping condition $|Dx-b|/|b|<10^{-9}$ 
for the force calculation and $10^{-14}$ for the Hamiltonian, which 
guarantees the reversibility of the molecular dynamics trajectories 
to high precision. 
The DDHMC algorithm for the up-down quark works efficiently
for $\kappa_{\rm ud}\le 0.13770$.

\begin{table}[t!]
\setlength{\tabcolsep}{6pt}
\renewcommand{\arraystretch}{1.1}
\centering
\caption{Summary of simulation parameters. Quark masses are
 perturbatively renormalized in the $\msbar$ scheme at the scale of $\mu=1/a$. 
The replay trick is applied for the case of $dH>2$. 
MD time is the number of trajectories multiplied by the trajectory
 length $\tau$. CPU time for unit $\tau$ is measured on 256 nodes of 
 the PACS-CS computer.}
\label{tab:param}
\begin{tabular}{lllllll | l}  \hline
$\kappa_{\rm ud}$ & 0.13700 & {0.13727} & 0.13754 & 0.13754 
 & {0.13770} & {0.13781} & 0.137785\\ 
$\kappa_{\rm s}$  & 0.13640 & {0.13640} & 0.13640 & 0.13660 
 & {0.13640} & {0.13640} & 0.13660\\  \hline 
HMC & DD & DD & DD & DD & DD & MP & MP2 \\
$\tau$   & 0.5 & 0.5 & 0.5 & 0.5 & 0.25 & 0.25 & 0.25\\
$(N_0,N_1,N_2,N_3,N_{4})$ &  (4,4,10) &  (4,4,14) &  (4,4,20) & (4,4,28) & (4,4,16) & (4,4,4,6) &(4,4,2,4,4)\\
                    &           &           &           &          &  & (4,4,6,6) & \\
$\rho_1$      & $-$ & $-$ & $-$ & $-$ & $-$ & 0.9995 & 0.9995\\
$\rho_2$      & $-$ & $-$ & $-$ & $-$ & $-$ & $-$ & 0.9990\\
$N_{\rm poly}$ & 180 & 180 & 180 & 220 & 180 & 200 & 220 \\
replay trick     & on  & on & on & on & on & off  & off \\
rate of $dH>2$ & 0\% & 0.08\% & 0.5\% & 0.1\% & 3\% & 2.8\% & 0.9\% \\
MD time & 2000 & 2000 & 2250 & 2000 & 2000 & 990 & 950\\
CPU time [hrs] & 0.29  & 0.44 & 1.3 & 1.1 & 2.7 & 7.1 & 6.0\\ 
$m_{\rm ud}^{\overline{MS}}$ [MeV] & 66.8(7) & 45.3(5) & 24.0(3) &
 21.0(3) & 12.3(2) & 3.5(2) & 3.5(1) \\
$m_{\pi}$ [MeV] & 702(7) & 570(6) & 411(4) & 385(4) & 296(3) & 156(2) & 164(4) \\ \hline
\end{tabular} 
\end{table}

As we reduce the up-down quark mass, the increasing fluctuations of 
the $||F_{\rm IR}||$ make the simulation unstable.
To suppress the fluctuations of $||F_{\rm IR}||$, 
we incorporate the mass preconditioning for the IR part (MPDDHMC),
which splits  the IR force $F_{\rm IR}$ into  $F_{\rm
IR}^{\prime}$ and  $\tilde{F}_{\rm IR}$ by introducing 
a new hopping parameter 
$\kappa^{\prime}_{\rm ud}=\rho_1\kappa_{\rm ud}$ with $\rho_1$ less than
unity. In the MPDDHMC algorithm we need four integers $(N_0, N_1, N_2,
N_3)$ to controll the four step sizes $\delta\tau_{\rm G},
\delta\tau_{\rm UV}, \delta\tau_{\rm IR}^{\prime}, \delta\tilde\tau_{\rm
{IR}}$. $N_2, N_3$ and $\rho_1$ are adjusted to reduce 
the fluctuations of $||F^{\prime}_{\rm IR}||$ and $||\tilde{F}_{\rm IR}||$.
We choose $\delta\tau_{\rm s}=\delta\tau_{\rm IR}^{\prime}$ 
for the strange quark force in the UVPHMC algorithm.

For the run at $\kappa_{\rm ud}=0.137785$ further 
mass preconditioning is applied to the shifted IR force $F^{\prime}_{\rm IR}$,
which is divided into $F_{\rm IR}^{\prime\prime}$ and 
$\tilde{F}^{\prime}_{\rm IR}$ using an additional hopping parameter 
$\kappa^{\prime\prime}_{\rm ud}=\rho_2\kappa^{\prime}_{\rm
ud}=\rho_2\rho_1\kappa_{\rm ud}$ with
$\rho_2$ less than unity.
We refer to this algorithm as MP2DDHMC because of two-level of 
mass preconditioning.  
In this case  five step sizes $\delta\tau_{\rm G}, \delta\tau_{\rm UV},
\delta\tau_{\rm IR}^{\prime\prime}, \delta\tilde\tau^{\prime}_{\rm IR},
\delta\tilde\tau_{\rm {IR}}$ are controlled by five integers $(N_0, N_1,
N_2, N_3, N_4)$.  
We adjust the values of $N_2, N_3, N_4$ and $\rho_1, \rho_2$ to keep stable
the fluctuations of $||F^{\prime\prime}_{\rm IR}||, ||\tilde{F}^{\prime}_{\rm IR}||, ||\tilde{F}_{\rm IR}||$. 
$\delta\tau_{\rm s}$ is equal to $\delta\tau_{\rm IR}^{\prime\prime}$.

For the MPDDHMC and the MP2DDHMC algorithms 
the inversion of the Wilson-Dirac operator on
the full lattice is composed of three steps. 
Firstly, we prepare the initial solutions employing 
the chronological guess with the last 16 solutions.
Secondly, we apply a nested BiCGStab solver consisting of 
the outer solver and the inner one. The latter with single precision 
arithmetic works as a preconditioner for the former operated 
with double precision.
We employ a stringent stopping condition 
$|Dx-b|/|b|<10^{-14}$ for the outer solver and 
an automatic tolerance control ranging from $10^{-3}$ to $10^{-6}$ 
for the inner solver.  
Thirdly, the nested BiCGStab solver is replaced by the GCRO-DR
(Generalized Conjugate Residual with implicit inner Orthogonalization
and Deflated Restarting) algorithm, once the inner BiCGStab solver
becomes stagnant during the inversion of the Wilson-Dirac operator.
   
\begin{figure}[b]
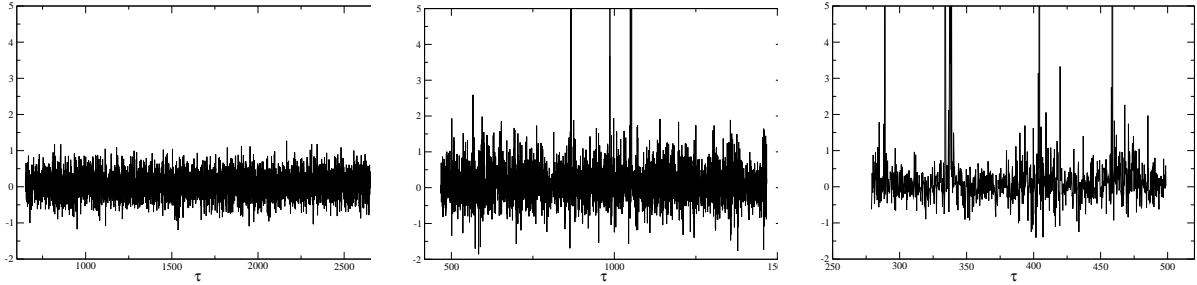

\vspace{3mm}
\begin{center}
\begin{tabular}{ccc}
\includegraphics[width=50mm,angle=0]{FIG/dH13727.eps}  &
\includegraphics[width=50mm,angle=0]{FIG/dH13770.eps}  &
\includegraphics[width=50mm,angle=0]{FIG/dH13781.eps}
\end{tabular}
\end{center}
\vspace{-.5cm}
\caption{$dH$ histories for $(\kappa_{\rm ud},\kappa_{\rm s})=$(0.13727,0.13640), (0.13770,0.13640) and (0.13781,0.13640) from the left.}
\label{fig:dH}
\end{figure}
\begin{figure}[b]
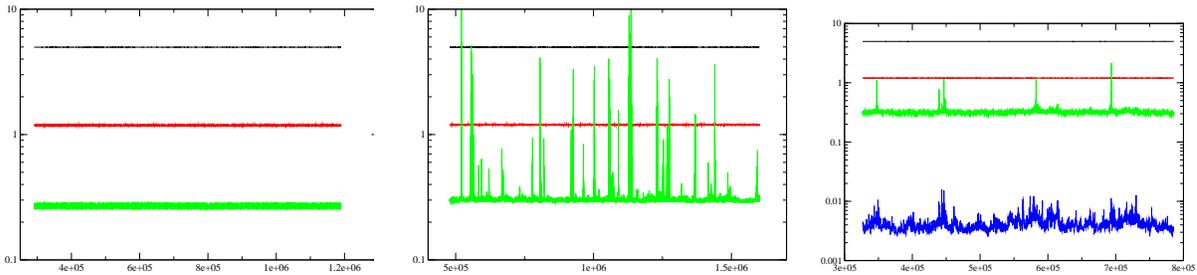

\vspace{3mm}
\begin{center}
\begin{tabular}{ccc}
\includegraphics[width=50mm,angle=0]{FIG/F13727.eps}  &
\includegraphics[width=50mm,angle=0]{FIG/F13770.eps}  &
\includegraphics[width=50mm,angle=0]{FIG/F13781.eps}
\end{tabular}
\end{center}
\vspace{-.5cm}
\caption{Force histories for $(\kappa_{\rm ud},\kappa_{\rm s})=$(0.13727,0.13640), (0.13770,0.13640) and (0.13781,0.13640) from the left.
In the left and middle figures black, red and green lines 
denote $F_{\rm G}$, $F_{\rm UV}$ and
$F_{\rm IR}$, respectively, with the DDHMC algorithm, 
In the right figure black, red, green and  blue lines are for 
$F_{\rm G}$, $F_{\rm UV}$, $F_{\rm IR}^{\prime}$ and $\tilde{F}_{\rm IR}$,
respectively, with the MPDDHMC algorithm.}
\label{fig:F}
\end{figure}

In Figs.~\ref{fig:dH} and \ref{fig:F} we show the $dH$ and the force 
histories at $\kappa_{\rm ud}=0.13727$ and 0.13770 with the DDHMC algorithm 
and those at $\kappa_{\rm ud}=0.13781$ with the MPDDHMC algorithm.
The time histories at $\kappa_{\rm ud}=0.13727$ are quite stable, whereas
the $\kappa_{\rm ud}=0.13770$ case shows the spike-like fluctuations 
of the IR force at a few \% rate of trajectries. 
For the $\kappa_{\rm ud}=0.13781$ run
we observe that the MPDDHMC algorithm succeeds in reducing the
fluctuations of the IR forces $F_{\rm IR}^{\prime}$ and $\tilde{F}_{\rm IR}$.

\section{Hadron spectrum} 
\begin{figure}[b]
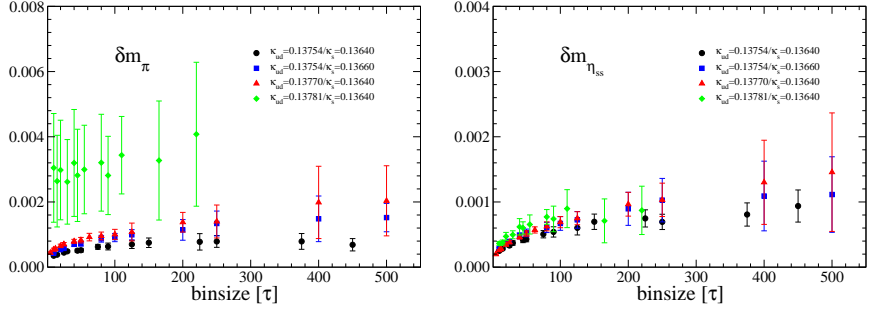

\vspace{3mm}
\begin{center}
\begin{tabular}{cc}
\includegraphics[width=55mm,angle=0]{FIG/binerr_pi.eps} &
\includegraphics[width=55mm,angle=0]{FIG/binerr_eta.eps}
\end{tabular}
\end{center}
\vspace{-.5cm}
\caption{Binsize dependence of the magnitude of error for $m_{\pi}$ (left) and $m_{\eta_{\rm ss}}$ (right)
 at $\kappa_{\rm ud} \ge 0.13754$}
\label{fig:binsize}
\end{figure}

We measure hadron correlators at ever 10 trajectories for 
$\kappa_{\rm ud}\le 13770$ and 20 trajectories for $\kappa_{\rm ud}\ge 13781$.
Light hadron masses are extracted from single exponential 
$\chi^2$ fits to the correlators
with an exponentially smeared source and a local sink.
In order to increase the statistics we take four source points with
different time slices for $\kappa_{\rm ud}\ge 0.13754$.
They are averaged on each configuration before the jackknife analysis. 
This reduces the statistical errors by typically 20--40\% for 
the vector meson and the baryon masses and less than 20\% 
for the pseudoscalar meson masses compared to a single source point.   
Statistical errors are estimated by the jackknife method. 
Figure~\ref{fig:binsize} shows the binsize dependence of 
the error for the pion mass and the ``$\eta_{ss}$'' meson mass. 
We observe that the magnitude of the error reaches
a plateau after 100--200 MD time. This feature seems almost independent of
the quark mass.
Since similar binsize dependences are found for other particle types, we
choose a binsize of 250 MD time at $\kappa_{\rm ud}<0.13770$.
At $\kappa_{\rm ud}=0.13781$ we employ a binsize of 110 MD time 
due to the lack of statistics.

\begin{figure}[b!]
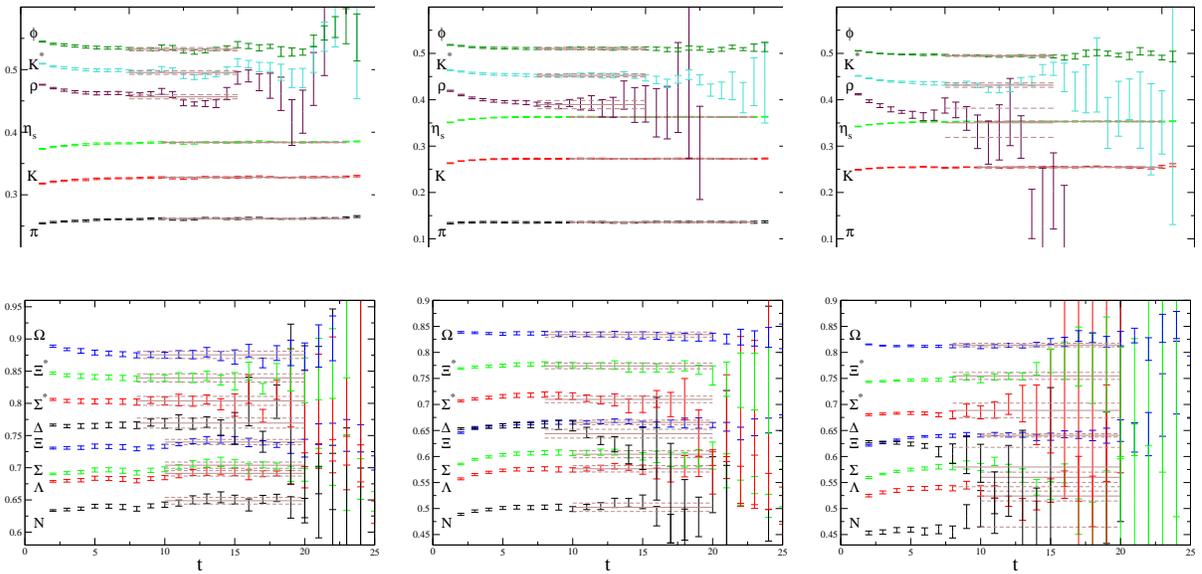

\vspace{3mm}
\begin{center}
\begin{tabular}{ccc}
\includegraphics[width=50mm,angle=0]{FIG/msn_13727.eps}  &
\includegraphics[width=50mm,angle=0]{FIG/msn_13770.eps}  &
\includegraphics[width=50mm,angle=0]{FIG/msn_13781.eps}  \\
\includegraphics[width=50mm,angle=0]{FIG/brn_13727.eps}  &
\includegraphics[width=50mm,angle=0]{FIG/brn_13770.eps}  &
\includegraphics[width=50mm,angle=0]{FIG/brn_13781.eps}
\end{tabular}
\end{center}
\vspace{-.5cm}
\caption{Effective masses for the mesons (top) and the baryons (bottom) for $(\kappa_{\rm ud},\kappa_{\rm s})=$(0.13727,0.13640), (0.13770,0.13640) and (0.13781,0.13640) from the left.}
\label{fig:Meff}
\end{figure}

\begin{figure}[t]
\vspace{3mm}
\begin{center}
\begin{tabular}{c}
\includegraphics[width=60mm,angle=0]{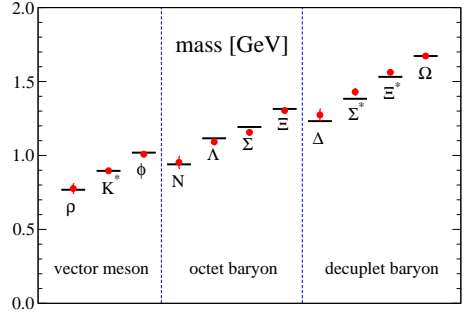}
\end{tabular}
\end{center}
\vspace{-.5cm}
\caption{Light hadron spectrum extrapolated to the physical point (red
 circles) in comparison with the experimental values (black bars).}
\label{fig:hyo}
\end{figure}

Figure~\ref{fig:Meff} shows the hadron effective masses  
at $\kappa_{\rm ud}=0.13727, 0.13770$ and 
$0.13781$. We observe clear plateau for the mesons except the $\rho$
meson at $\kappa_{\rm ud}=0.13781$ and also good signal 
for the baryons thanks to a large volume.
Especially, the $\Omega$ baryon has a stable signal 
and a weak up-down quark mass dependence for our simulation parameters.   
Taking advantage of this virtue we choose the 
$\Omega$ baryon as input to determine the lattice cutoff.
Combined with the additional inputs of $m_\pi$ and $m_K$ to determine
the physical up-down and strange quark masses,
we obtain $a^{-1}=2.176(31)$ GeV.
In this procedure we employ the SU(2) ChPT analyses 
for the quark mass dependences
of $m_\pi$, $m_K$, $f_\pi$ and $f_K$ taking account of the finite 
size corrections evaluated at the one-loop level\cite{pacscs,kadoh}. 
For $m_\Omega$ we assume the linear quark mass depenedences. 
With the use of this cutoff we find that the lightest
pseudoscalar meson mass we have reached is about 160 MeV.
To obtain the vector meson masses and the baryon masses
at the physical point
we avoid the chiral analyses based on the heavy meson effective theory 
or the heavy baryon ChPT
because of their poor convergences in the chiral expansions.
We instead use linear chiral extraporations to  the physical point.
In Fig.~\ref{fig:hyo} we compare the light hadron spectrum 
at the physical point with the experimental values. 
The largest discrepancy is at most 3\%, albeit errors are still not small for the $\rho$ meson, the nucleon and the $\Delta$ baryon.
It should be also noted that our results contain 
possible $O((a\Lambda_{\rm QCD})^2)$ cutoff errors.

We calculate the bare quark masses using the axial vector
Ward-Takahashi identity (AWI) defined by
$am^{\rm AWI}=\lim_{t\rightarrow \infty}
{\langle \nabla_4A_4^{\rm imp}(t)P(0) \rangle}/(2\langle P(t)P(0)\rangle)$
where $A_4^{\rm imp}$ is the nonperturbatively 
$O(a)$-improved axial vector current\cite{A4I}.
Employing the perturbative renormalization factors $Z_A$ and $Z_P$ 
evaluated up to one-loop level \cite{Z1, Z2}, we obtain
\begin{eqnarray}
m_{\rm ud}^{\overline{\rm MS}}(\mu=2{\rm GeV})=2.527(47){\rm MeV},&&
m_{\rm s}^{\overline{\rm MS}}(\mu=2{\rm GeV})=72.72(78){\rm MeV}.
\end{eqnarray}
The physical up-down quark mass is 30\% smaller than 
our lightest one $m_{\rm ud}^{\overline{\rm MS}}(\mu=1/a)=3.5$ MeV 
at $(\kappa_{\rm ud}, \kappa_{\rm s})=(0.13781, 0.13640)$.
The results for the pseudoscalar meson decay constants are given by
\begin{eqnarray}
f_\pi=134.0(4.2){\rm MeV}, \ \ & f_K=159.4(3.1){\rm MeV},  \ \ & f_K/f_{\pi}=1.189(20)
\end{eqnarray}
at the physical point with the perturbative $Z_A$.
They are consistent with the experimental values within the errors.
Our concern about the values for the quark masses and the pseudoscalar 
meson decay constants is the use of the perturbative renormalization factors
which might cause sizable systematic errors. 
We are now calculating the nonperturbative $Z_A$ and $Z_P$ with the
Schr{\"o}dinger functional scheme.

\begin{acknowledgments}
Numerical calculations for the present work have been carried out
on the PACS-CS computer 
at Center for Computational Sciences, University of Tsukuba. 
A part of the code development has been carried out on Hitachi SR11000 
at Information Media Center of Hiroshima University. 
This work is supported in part by Grants-in-Aid for Scientific Research
from the Ministry of Education, Culture, Sports, Science and Technology
(Nos.
16740147,   
17340066,   
18104005,   
18540250,   
18740130,   
19740134,   
20340047,   
20540248,   
20740123,   
20740139    
).
\end{acknowledgments}


\end{document}